\documentclass[11pt]{article} 
\usepackage{graphics}
\textheight 8.in \title{THERMODYNAMICS WITH 3 SPINS}
\author{EDWARD MAY \\may @ anl.gov \\ \and JACK L. URETSKY
\\jlu@anl.gov \\ High Energy Physics Division, Argonne
National Laboratory} \date{\today}                          
\begin{document}
\setlength{\unitlength}{.7mm} \hfill ANL-HEP-PR-09-11 \today
\maketitle \section{ABSTRACT} Glauber dynamics, applied to
the one-dimensional Ising model, provides a tractable model
for the study of non-equilibrium, many-body processes driven
by a heat bath.  We explain Glauber`s dynamical Ising model
in the context of a system comprising three ``spins'' in a
1-dimensional Ising model.  We first present the model
without a magnetic field and then, following Glauber, add an
oscillating magnetic field.  We then show that Glauber's
demonstration of the fluctuation-dissipation theorem carries
over to the 3-spin case. A web-link to a simulation code and
an interactive video display of the ``spins'' flipping from
a non-equilibrium to an equilibrium configuration (zero
magnetic field) is included. \vspace{10mm}
\section{INTRODUCTION} The study of thermodynamics
traditionally deals with equilibrium properties of matter. 
However, as remarked by Le Bellac, {\em et al.}(\cite{Le
Bel},p. 335) ``this is rather restrictive since
non-equilibrium phenomena, such as heat conduction or
diffusion, are of great interest and cannot be ignored.'' 
On the other hand, our understanding of the properties of
out-of-equilibrium matter is still something of a work in
progress, as one might conclude from Kubo's brief summary of
modern developments in the theory of irreversible processes.
(\cite{Kubo}, p. 374).

	The Ising model \cite{Lenz} considers arrays of coupled
two-component spins, and is often discussed in connection
with ferromagnetism in modern textbooks.  The model lived
most of its life as an equilibrium model, until, as
discussed by R\`{a}cz \cite{ZR}, Glauber\cite{RJG}initiated
an industry of kinetic spin models (currently, "Glauber
Ising" gets about 1.4 million hits on Google) by providing 
for single spin-flips induced by a heat bath followed by a
relaxation to equilibrium.  A solution for an arbitrary
number of spins was subsequently given by Felderhof
\cite{Felderhof} .  Kawasaki quickly followed Glauber's
paper with a spin-exchange model that conserved total spins.
 Garrido, {\em et al.} \cite{Garr}, using Glauber dynamics
with competing heat baths, demonstrated the existence of
non-equilibrium phase transitions in 1-dimension with
short-range forces.  Lage \cite{Lage} extended the use of
Glauber dynamics to the Potts model, taking care to point
out that the extension is not unique, that is, that there
are multiple possible dynamics that could be added to the
original Potts model.

	Many authors have recognized that the Ising model can
also serve as a lattice-gas model, the two ``spin''
orientations being considered as lattice occupation numbers,
$0$ and $1$.  Chomaz {\em et al.}\cite{chom} invoke such an
interpretation to discuss the formation and evolution of
compact stars. The Ising model has also found application to
the discussion of the thermodynamic properties of
Schwarzschild black holes \cite{kas} A recent paper by
Mazilu and Williams \cite{Maz} "introduces nonequilibrium
statistical mechanics" by solving exactly a two-temperature
(1-d) Ising model with four spins.

	The Ising Hamiltonian for a lattice with $N$ spins is
\cite{Huang} \begin{equation} {\cal  H} =-J \sum_{i = 1}^{N}
s_{i}s_{i+1} -H\sum _{i=1}^{N}s_{i}         \label{Ham}
\end{equation} with index $N+1 \equiv 1$.  $H$ is the
strength of an applied external magnetic field, which may be
time dependent.  The coupling strength $J$ is positive for
an attractive interaction that would simulate ferromagnetism
if all the spins had the same value, thereby indicating
alignment.

	Statistical mechanics typically deals with very large
systems, see {\em e.g.} \cite{Huang}.   Glauber's \cite{RJG}
paper accordingly considers the infinite 1-d Ising chain (as
well as the single spin case).  Chemists and biologists,
however, are often interested in very small systems, such as
molecules with just a few binding sites.   Ter Bush and
Thompson \cite{TBT} worked out the Glauber dynamics of
2-spin and 4-spin Ising chains in connection with a model
for the functioning of hemoglobin.

	\section{GLAUBER-ISING WITH 3 SPINS} The 3-spin 1-d
dynamic Ising model displays many of the features of larger
finite Ising systems and is easily handled by elementary
methods in the case of zero magnetic field.  It is assumed
that each individual spin, $s_{i}$ flips randomly between
the values $+1,  -1$ with a probability per unit time
(measured in arbitrary time units) that depends upon the
orientations of the adjacent spins according to
\begin{equation} w_{i}(s_{i})=[1-\frac{1}{2}\gamma s
_{i}(s_{i-1}+s_{i+1})]      \label{eq:g1} \end{equation} The
constant $\gamma $ is related to the spin-coupling $J$ and
the temperature of a presumed heat bath, that drives the
random orientation transitions of the individual spins, by
\cite{RJG} \begin{equation} \gamma =
\coth\left(\frac{2J}{kT}\right)     \label{gam}
\end{equation} where T is the temperature of the heat bath.
Appendix A includes a table of the values of the $w_{i}$'s.

	Designating the probability of a given spin
configuration by $p(s_{1},s_{2},s_{3};t)$, where each
$s_{i}$ can take  either of the values $+,-$,  the time
dependence of the probabilities is given by the master
equation \begin{equation}
\frac{dp(s_{1},s_{2},s_{3};t)}{dt}=-\sum_{i} w_{i}(s_{i})
p(s_{1},s_{2},s_{3};t)+ \sum_{j}w_j(-s_{j})p(s_{j-1},
-s_{j},s_{j+1};t)                                         
\label{eq:gl2} \end{equation} with the convention that 
subscripts $0$ and $4$ are respectively the same as $       
     3$ and $1$. As pointed out in connection with Eq. 17 of
\cite{RJG}, the coupling strength $\gamma $ is limited to
the range $[+1,-1]$, the extreme values corresponding to
zero temperature of a presumed heat bath, and the negative
value corresponding to a repulsive interaction between
adjacent spins.  The coupling vanishes in the limit of
infinite temperature of the heat bath.

	Now let $V_{+ } (V_{-})$ represent the probability for
the spins all having the value $+1 (-1)$, and $X_{+}(X_{-})
$ the sum of the probabilities that  exactly two spins have
the values $+1(-1)$.  The individual probabilities for the
latter configuration are denoted by $x_{\pm i}$ with the
$``i''$ denoting the position of the odd spin.  Also, define
$V= V_{+} +V_{-}$,  $X= X_{+} +X_{-}$ , $U= V_{+} -V_{-}$,
$Y= X_{+} -X_{-}$, $x_{i}= x_{+i} +x_{-i }$, and  $y_{i}=
x_{+i} -x_{-i}$.  Clearly, in the case of three spins
\begin{equation} V+X=1 \label {eqs1} \end{equation}

	\section{ZERO MAGNETIC FIELD} Define a vector
\begin{equation} \Psi = \left[ \begin{array}{cccc} V   \\ X 
 \\ U  \\ Y \end{array}  \right] \label{eq:psi}
\end{equation} Then, In Glauber's ``alternative'' approach,
the master equation may be written as (the dot denotes time
differentiation) \begin{equation} \frac{d}{dt} \Psi  \equiv 
\dot{\Psi } =M\Psi \label {eq:Mas} \end{equation} where the
matrix $M$ is

	$ \left[  \begin{array}{cccc} -3(1-\gamma )& (1+\gamma 
) & 0  &  0     \\ 3(1-\gamma )& -(1+\gamma  ) & 0 & 0    
\\ 0  & 0     & -3(1-\gamma )& (1+\gamma  )    \\ 0 &  0    
& 3(1-\gamma )& -(5+\gamma ) \end{array} \right]  $

	The eigenvalues of the matrix M are $0, -2(1-\gamma ),
-2(2-\gamma ), -6$.  The zero eigenvalue corresponds to
equilibrium.  There is a ``parity'' symmetry in the
equations of motion, corresponding to the simultaneous
flipping of all three spins, so that both U and Y must be
equal to zero at equilibrium, and the equilibrium solution
is $V=\frac{(1+\gamma )}{2(2-\gamma )}$, as would be
expected from the static Ising model.  It is evident that
the six configurations with one odd spin must be equally
populated.

	The solutions are \begin{eqnarray} X(t) & =  &
X(0)e^{-2(2-\gamma )t} + \frac{3(1-\gamma )}{2(2- \gamma)
}[1-e^{-2(2-\gamma )t}]  \nonumber   \\ U(t) & = 
&\frac{1}{2(2+\gamma)} \{U(0)[(1-\gamma
)e^{-6t}+3(1+\gamma)e^{-2(1-\gamma )t}-[(1+ \gamma
)Y(0)[e^{-6t}-e^{-2(1-\gamma )t}]\} \nonumber  \\ Y(t) & = 
&\frac{1}{2(2+\gamma )} \{Y(0)[3(1+\gamma )e^{-6t}+(1-\gamma
)e^{-2(1-\gamma )t}]-3(1- \gamma
)U(0)[e^{-6t}-e^{-2(1-\gamma )t}] \} \label{eq:UY2}
\end{eqnarray} The first of these equations provides for
relaxation to the classical Ising equilibrium solution, as
shown in the accompanying Figure, where, in the initial
configuration the three spins are aligned in the positive
direction ($X(0)=0, U(0)=+1$, so that $Y(0)=0$).  The scale
of the $t-$variable is arbitrary
\includegraphics{theory.eps}

	\subsection{Average Values} Glauber also suggests that
in many situations the relevant observable quantities might
be average values and correlations of the individual spins,
rather than the detailed configurations.  Accordingly, he
defines the single spin average value to be, for the $k'th$
spin \begin{equation} q_{k}(t)  \equiv  <s_{k}>= \sum_{s }
s_{k}p(s_{1},s_{2},s_{3};t)         \label{q:def}
\end{equation} where the sum is over the eight possible
values of the three spins.  The time dependence of the
$q_{k}$ is given by \begin{equation} \dot{q_{k}}(t) = 
-[q_{k}(t) -\frac{\gamma }{2}[q_{k-1}(t)+q_{k+1}(t)] ]      
   \label{eq:qtime} \end{equation} The solution to Eq.
\ref{eq:qtime}  is (see Appendix B ) \begin{equation}
q_{k}(t) = \frac{1}{3}[q_{k}(0)(e^{-(1-\gamma )t} +
2e^{-(1+\frac{\gamma }{2})t}] +[q_{(k-1)}(0)+ 
q_{k+1}(0)][e^{-(1-\gamma )t} -e^{-(1-\frac{\gamma }{2})t}] 
                             \label{eq:q} \end{equation}

	Evidently. at equilibrium. reached from any starting
configuration, the average value of each spin is zero, as
might be expected.

	\subsection{2-spin correlations} The 2-spin correlation
function is defined \cite{RJG}, in the 3-spin case, by
\begin{equation} r_{j,k}(t)  \equiv  <s_{j}s_{k}> =
\sum_{s}s_{j}s_{k}p(s_{1},s_{2},s_{3};t)   \label{rjk}
\end{equation} which guarantees that $r_{j,j} \equiv 1$.  
The $r's$ are, of course, symmetric in the two indices.  
The time dependences are given by \begin{eqnarray}
\dot{r}_{i,i+1}&=&-2r_{i,i+1} + \frac{\gamma
}{2}[r_{i,i-1}+r_{i-1,i+1}] +\gamma \nonumber   \\
\dot{r}_{i,i-1}&=&-2r_{i,i-1} + \frac{\gamma
}{2}[r_{i,i+1}+r_{i-1,i+1}] +\gamma  \nonumber   \\
\dot{r}_{i-1,i+1}&=&-2r_{i-1,i+1} + \frac{\gamma
}{2}[r_{i,i-1}+r_{i,i+1}] +\gamma    \label{eq:rtime}
\end{eqnarray} The solutions are (see Appendix C)
\begin{equation} r_{i,j}(t) =
r_{i,j}(0)e^{-\frac{1}{2}(4+\gamma )t} +
\frac{1}{3}R(0)[e^{-(2-\gamma )t}-e^{-(\frac{1}{2}(4+\gamma
)t}] + \frac{\gamma }{(2-\gamma )}[1-e^{- (2-\gamma )t}]    
  \label{rt} \end{equation} where $R$ is the sum of the
three independent correlation functions.  The final term of
Eq \ref{rt} indicates that the 3 spins are partially
correlated at all times. \vspace{1in}

	\section{MAGNETIC FIELD} Glauber observes that a dynamic
Ising model with a magnetic field H will have the same
equilibrium solution as the static Ising model if $w_{i}$ in
Eq \ref{eq:g1} is replaced by \begin{equation}
w_{i}'(s_{i})=w_{i}(s_{i})[1-\beta s_{i}]         
\label{g2} \end{equation} where \begin{equation} \beta
\equiv  \tanh (\frac{\mu H}{kT})          \label{eq:bet}
\end{equation} and $\mu $ is the magnetic moment associated
with one of the  spins. The Matrix M of the previous section
is then replaced by \begin{equation} \left[
\begin{array}{cccc} -3(1-\gamma )& (1+\gamma  ) & 3(1-\gamma
)\beta & (1+\gamma )\beta    \\ 3(1-\gamma )& -(1+\gamma  )
& -3(1-\gamma )\beta & -(1+\gamma )\beta    \\ 3(1-\gamma
)\beta &  (1+\gamma )\beta     & -3(1-\gamma )& (1+\gamma  )
   \\ -3(1- \gamma )\beta & (3- \gamma )\beta     & 
3(1-\gamma )& -(5+\gamma ) \end{array} \right]       
\label{eq:mag} \end{equation} \subsection{Zero Temperature:
A Phase Transformation?} The zero temperature ( $\gamma
=\beta =1$) eigenvalues are readily found to be \vspace{1in}
$0$ and $4$, each occurring twice.  The solutions may be
written as \begin{eqnarray} X(t) & = & \{X(0) +  2t[X)0)
-Y(0)]\}e^{-4t}     \nonumber   \\ Y(t) & = & \{Y(0) + 
2t[X)0) -Y(0)]\}e^{-4t}     \nonumber   \\ U(t) & = & \{U(0)
+ X(0)(1-e^{-4t}) -  2t[X(0) -Y(0)]\}e^{-4t}      \nonumber 
 \\ V(t) & = & 1- X(t)                                      
\label{eq:zersol} \end{eqnarray}

	The equilibrium solution, which depends upon the initial
conditions, has no counterpart in the original Ising model,
where the concept of ``initial'' plays no role.  The system
ultimately relaxes to the configuration \begin{eqnarray}
V(t) &  \sim & 1           \nonumber  \\ U(t) & \sim  & U(0)
+ X(0)    \nonumber   \\ &\equiv & 1- 2V_{-}(0)   
\label{eq:asym} \end{eqnarray} \vspace{1in} where the last
identity results from the fact that there are only three
spins.

	A number of authors have treated the zero temperature
limit of an Ising system as a phase boundary (see,
generally, Privman in \cite{ZR}).  That this is a tempting
analogy may be seen from the fact that  in the zero
temperature limit, the parameter $|\beta |$ is $1$ when a
magnetic field, no matter how small, is present, and $0$ if
the field is absent.   This parameter breaks the ``parity''
symmetry of the  equilibrium solutions, resulting in the
non-zero asymptotic values of $Y(t)$ and $U(t)$.
\vspace{1in}

	\subsection{Magnetic Field at Low, but Non-Zero
Temperature} The eigenvalues of the Master Equation matrix,
Eq. \ref{eq:mag}, must, in general, be determined from a
cubic equation.  I have investigated the solutions for the
low, but non-zero, temperature case where $\gamma =\beta
=0.90$.  I find the 3 non-zero eigenvalues to be
approximately $-4.81, -3.05,$ and $  -0.538$.  These may be
compared with the values $-6, -2.2. -0.2$,  in the absence
of a magnetic field ($\beta =0$), and  $-4, -3.8, -.6  [-4,
-2(1+\gamma ),-6(1-\gamma )]$ in the presence of a very
large magnetic field ($\beta =1$).  The magnetic field
obviously breaks the parity symmetry, favoring $+$-spins
over $-$-spins (for positive $\mu H$).

	The secular equation for the eigenvalues of the Master
Equation matrix involve the parameter $\beta$ to at least
the second power, so that to linear order in the magnetic
field the eigenvalues are unchanged from their zero-field
values.  To linear order in $\beta $ the relaxation rates
are just the zero-field rates. The solutions to the same
order in $\beta$, for a time dependent external field $\beta
=\beta _{0}e^{-i\omega t}$ are:

	\begin{eqnarray} \Delta U & = &  
-\frac{\beta_{0}}{2+\gamma } \langle [X(0)-3\frac{1-\gamma
)}{2(2- \gamma )}] \frac{4\gamma    (1+\gamma )}{2+i\omega
}(e^{-[2(2-\gamma )+i\omega ]t}- e^{-2(1-\gamma )t})+
\nonumber \\ &     &  \frac{2(2-\gamma }{2(1+\gamma -i\omega
)} (e^{-[2(2-\gamma )+i\omega ]t}-e^{-6t}) -\frac{3(1-\gamma
^{2})}{(2(1-\gamma -i\omega )}\frac{2+\gamma }{2-\gamma
}(e^{-i\omega t} -e^{-2(1-\gamma )t} \rangle \nonumber \\
\Delta Y & = &  -\frac{\beta_{0}}{2+\gamma } \langle 
[X(0)-3\frac{(1-\gamma )}{2(2- \gamma )}]\frac{4\gamma   
(1-\gamma )}{2+i\omega }(e^{-[2(2-\gamma )+i\omega ]t}-
e^{-2(1-\gamma )t})- \nonumber  \\ &    &   \frac{6(2-\gamma
)}{2(1+\gamma -i\omega )}(e^{-[2(2-\gamma) +i\omega
]t}-e^{-6t}) -\frac{3(1-\gamma )^{2}}{2(1-\gamma)-i\omega
]}\frac{2+\gamma }{2-\gamma }( e^{-i\omega t}-e^{-2(1-\gamma
)t} ) \rangle     \label{eq:UYmag} \end{eqnarray}

	The equilibrium values are given by \begin{eqnarray}
U(t)  &  \sim  \frac{12\beta_{0}(1-\gamma )}{(4-\gamma
^{2})[2(2-\gamma )-i\omega ]}e^{-i\omega i}  \nonumber \\
Y(t)   &  \sim \frac{12\beta _{0}(1-\gamma )^{2}}{(4-\gamma
^{2})[2(2-\gamma )+i    \omega ]}e^{-i\omega i}  
\label{eq:UY} \end{eqnarray} Setting $\omega = 0$ makes it
obvious that the magnetic field breaks the up-down symmetry
of the no-field  solutions.

	\subsection{Average Values with a Weak Magnetic Field}
In place of Eq \ref{eq:qtime} the average value of spin k
must satisfy,  in the presence of a (weak) magnetic field
$H=H_{0}\exp^{i\omega t}$ \begin{equation} \dot{q_{k}} =
-q_{k}+\frac{\gamma }{2}(q_{k-1}+q_{k+1})+\beta
[1-\frac{\gamma }{2}(r_{k-1,k}+r_{k,k+1})] \label{Eq:qmag}
\end{equation} We suppose that the system is close to
equilibrium, and approximate by using the equilibrium value
$\frac{\gamma }{2-\gamma }$ of the correlation functions
$r_{i,k}$.  The solution, after the transients have died
off, is \begin{equation} q_{k}(t) \sim 3\beta
_{0}(\frac{2+\gamma }{2-\gamma })\frac{1-\gamma }{1-\gamma
-i\omega }e^{-i\omega t}        \label{qkmag} \end{equation}
The average value of Glauber's stochastic magnetization'', 
$<M(t)> \equiv \mu \sum_{k}q_{k}(t)$,  is then
\begin{equation} <M(t)>= 3\frac{\mu
^{2}H}{kT}(\frac{2+\gamma }{2-\gamma })\frac{1-\gamma
}{1-\gamma -i\omega }e^{-i\omega t}                
\label{eq:M} \end{equation} The time-delayed correlation
function $<M(0)M(t)>$ (see Appendix D)is \begin{equation}
<M(0)M(t)> = 3\mu ^{2}(\frac{1+\frac{\gamma
}{2}}{1-\frac{\gamma }{2}})e^{-(1-\gamma )|t|}
\label{eq:tcor} \end{equation} The frequency-dependent
susceptibility $\chi (\omega )$, defined by $<M(t)> =\chi
(\omega )H$ is, then \begin{equation} \chi  (\omega )=
3\frac{\mu ^{2}}{kT}\frac{2+\gamma }{2-\gamma
}\frac{1-\gamma }{1-\gamma -i\omega }     \label{eq:chi}
\end{equation}

	\subsection{The Fluctuation-Dissipation Theorem} The
fluctuation-dissipation theorem, see {\em e.g.}, \cite{Rubi}
was probably motivated by Johnson's discovery of electrical
"noise"  voltage across  otherwise isolated resistors
\cite{Johnson}.

	Nyquist \cite{Nyquist} subsequently related Johnson's
voltage to "thermal agitation" of charges in the resistor,
and predicted the magnitude of the voltage $V$ in a
frequency interval $d\nu $ to depend upon the values of the
ambient temperature $T$ and the resistance $R$, according to
\begin{equation} V^{2} d\nu = 4RkTd\nu     \label{eq:Nyq}
\end{equation} \nopagebreak In the present context,
following Glauber, we can let the magnetic field frequency
$\omega$ simulate the frequency of the ``thermal
agitation''.  Then the Fourier transform of the
magnetization correlation function, Eq \ref{eq:tcor} is
\begin{equation} \int
_{-\infty}^{\infty}dt<M(0)M(t)>e^{i\omega t} = 3\mu
^{2}(\frac{1+\frac{\gamma }{2}}{1-\frac{\gamma
}{2}})\frac{2(1-\gamma )}{(1-\gamma )^{2}+\omega
^{2}}=\frac{2kT}{\omega }Im\chi (\omega)      
\label{eq:fudis} \end{equation} Thus $\frac{2kT}{\omega }$
times the imaginary, or absorptive, part of the
susceptibility, in this magnetic example, plays the role of
the resistance  in the case of Johnson noise.  As Glauber
points out, relations such as this may be useful in finding
the effect of a weak field upon general functions of the
spin variables in Ising-type models.

	\section{Discussion and Conclusions} The dynamic Ising
model in one dimension would seem, at first blush, to be
somewhat insipid---it can do nothing except relax to
equilibrium.  The paths to equilibrium, however, can involve
all eight of  the possible configurations, in the case of
three spins.  This raises the possibility of  interesting
patterns occurring in similar systems with larger numbers of
spins.  Even the three-spin model becomes more interesting,
when used interactively, that is, by repeatedly shifting the
equilibrium temperature.  This is done by repeatedly
changing the value of the $\gamma $ parameter, randomly or
otherwise.

	The three-spin model with a weak magnetic field shows an
amusing relationship to the infinite-spin model considered
by Glauber.  The susceptiblity in Glauber?s Eq. 94 (see also
his Eq. 56) contains a term $\frac{1+\eta }{1- \eta }$,
which can be re-expressed as $\sqrt{\frac{1+\gamma
}{1-\gamma }}$. The equivalent three-spin term in Eq
\ref{eq:chi} is just the small $\gamma $ (weak-coupling)
approximation to Glauber's expression.

	{\em Acknowledgements}  I am deeply indebted to Dr.
Jason Green for introducing me to the notion of
thermodynamics of small systems, to Cosmas Zachos for
bringing Glauber's paper to my attention, and to both of
them for many pleasant and fruit ful dicussions, and Dr.
Robert Blair for assistance with the manuscript. This work
was supported in part by the U.S. Department of Energy,
Division of High Energy Physics, under Contract
DE-AC02-06CH11357

	\appendix{APPENDICES} \section{TRANSITION PROBABILITIES}
\begin{center}

	$\begin{array}{lllccc} &                      &         
                &          w_{i}s_{i}      &                
                  &                          \\ s_{1}	   
&	   s_{2}       &	   s_{3}       &	    w_{1}s_{1} 
  &	   w_{2}(s_{2}    & 	   w_{3}s_{3}   \\ +1        
&      +1		&	      +1	Ê&	1-\gamma		&  
1-\gamma		&	1-\gamma	        \\+1	     &     
+1		&	       -1	 &	1			&	1       
 		&	1+\gamma       \\ +1	     &	    
-1		&	      +1	 &	1			&  
1+\gamma		&		1	        \\ +1	     &      
-1		&	       -1	
&	1+\gamma	&	1			&		1	        \\
-1           &      +1		&	      +1	
&	1+\gamma	&	1			&		1	        \\
-1           &      +1 	&	       -1	
&	1			&   1+\gamma		&		1	       
\\ -1           &       -1		&	      +1	
&	1			&	1		         &	1+\gamma      
\\ -1           &       -1		&	       -1	 &     
1-\gamma		&   1-\gamma		&	1-\gamma
\end{array}$ \end{center}

	\section{AVERAGE VALUE SOLUTIONS} Summing Eq.
(\ref{eq:qtime}) over k (for arbitrary number of spins N),
and defining $Q(t)\equiv \sum _{k}q_{k}(t)$ gives the
equation \begin{equation} \dot{Q(t)}= -(1-\gamma )Q(t)   
\label{eq:Qdot} \end{equation} with the solution
\begin{equation} Q(t)=Q(0)e^{-(1-\gamma)t}  \label{eq:Q}
\end{equation} Then, in the present instance of 3 spins we
can rewrite Eq (\ref{eq:qtime} (\cite{TBT} Eq 32 {\em et
seq}) \begin{equation} \dot{q}_{k}(t)= -(1+\frac{\gamma
}{2})q_{k}(t) +\frac{\gamma }{2}Q(t) \end{equation} which
has the solution \begin{equation}
q_{k}(t)=e^{-(1+\frac{\gamma }{2})t}[q_{k}(0) +\frac{\gamma
}{2}Q(0)\int_{0}^{\tau }e^{\frac{3\gamma }{2}\tau }d\tau ]
\end{equation} which, in turn, gives Eq \ref{eq:q}.

	\section{AVERAGE CORRELATIION FUNCTION SOLUTION} In
keeping with the approach in Appendix B, we can define a
quantity \begin{equation} R=\sum_{i<k=1}^{3}r_{i,k}    
\label{defR} \end{equation} The sum of the three equations
(\ref{eq:rtime}) then gives \begin{equation} \dot{R}-
-(2-\gamma )R+3\gamma \end{equation} from which we deduce
that \begin{equation} R(t) = R(0)e^{-(2-\gamma )t} +
\frac{3\gamma }{2-\gamma }[1-e^{-(2-\gamma )t}]
\end{equation} but each of the equations (\ref{eq:rtime})may
be written as \begin{equation} \dot{r}_{i,k}=\frac{\gamma
}{2}(R-r_{i,k}) +\gamma \label{rik} \end{equation} from
which the solution given by Eq \ref{rt} may readily be
obtained by standard methods.

	\section{TIME DELAYED CORRELATION AND MAGNETIZATION} The
average two-spin time-delay correlation function
\begin{equation}
<s_{j}(0)s_{k}(t)>=r_{i,j}(0)\exp^{-(1-\gamma )t}
\end{equation}  is

	\begin{eqnarray} \lefteqn{<s_{j}(0)s_{k}(t)> =} 
\nonumber  \\ & &\frac{1}{3}<s_{j}(0)\{
s_{k}(0)[2\exp^{-(1+\frac{\gamma }{2})t}+\exp^{-(1-\gamma
)t}]  \nonumber \\ &
&[s_{k-1}(0)+s_{k+1}(0)]\}>[\exp^{-(1-\gamma
)t}-\exp{-(1+\frac{\gamma }{2})t}]\} =   \nonumber \\ &
&\frac{1}{3}[r_{j,k}(0)(2\exp^{-(1+\frac{\gamma
}{2})t}+exp^{-(1-\gamma )t}-[r_{j,k-1}(0)+r_{j,k+1}(0)]
\nonumber \\ & &\;\;(\exp^{-(1+\frac{\gamma
}{2})t}-exp^{-(1-\gamma)t}) = r_{i,j}(0)\exp^{-(1-\gamma )t}
 \label{eq:tdelay} \end{eqnarray}: where the last equality
reflects the fact that the $r_{j,k}$'s  are all equal at
equilibrium.  Summing over both indices (remembering that
$r_{j,j}=1$) and multiplying by $\mu ^{2}$ then leads
directly to Eq. \ref{eq:tcor}

	\section{SIMULATION AND VIDEO DISPLAY} There are eight
possible configurations of the three two-valued spins.  We
number these as states $N=1,..,8$ according two the scheme:
\\N = $1 (8)$ for configurations $V_{+} (V_{-})$ 
\\$N = 1+i$ for configurations $x_{+i}$ 
\\$N=3+i$ for configurations $x_{-i}$. \\The eight states are shown in eight .jpg files.

Transitions among the states, starting from some arbitrarily chosen initial state, with $\gamma $set in the range [-1,1],
are encoded in an interactive python program which may be 
downloaded from http://www.hep.anl.gov/may/jlu/ThermoWith3spins/index.html. 
Just download and open the .tar file at that location and
see the resulting README file for more information.

	The program simulates the relaxation to equilibrium of a
3-element Ising chain, and is reproduced at the end of this Appendix, augmented with a graphics display.  Programs, files and other
supporting information can be found at the Web URL address cited above.
The Web site also presents results of simulation
runs as a graphical comparison of the simulation results
with the a corresponding solution of the Master Equation.

	The graphical display of a run shows three arrows making transitions
among the eight possible states. Each transition is
one that is accomplished with a single-spin flip, as
illustrated in the ''cube`` file. also available
at the URL cited above. The transition rates, given in
Appendix A, are assumed to be the rates of Poisson Processes
\cite{FellerIPois}, where the values in Appendix A are
multiples of a basic rate ``r'' specified in The body of the
program.

	The program starts with the python loop command ''while
k  <= numberOfIter''. As presently set up, this loop is
first entered with the spins in the ''ferromagnetic`` state
1'', displayed according the the variable ``back1''.  Prior
to the loop we define functions including eight functions
denoted by ``$A_{n}$''  (n=1..8), and a function
``exponen(x)''.  Each $A_n$ function is called when the
system is in the state corresponding to the value of ``n''
and returns three parameters on exit, namely, which of the
three possible transitions will be made to a successor
state, the rate at which the transition will be made, and
the time spent in the current state prior to the transition.

	The Poisson flip rate for the successor state is given
by the parameter ``x'', specified in calling the exponen(x) function. 
That function randomly determines a waiting time z in the successor
state in accordance with the exponential distribution of
waiting times for the first event of a Poisson process.  The probability, p(z), that the first event has occurred after elapsed time z 
\begin{equation}
 p(z) = \exp(-zx) 
\end{equation}
 where
$p(z)$ is a random number in the range $[0,1]$ and $z$ is
the elapsed time variable (in the range $[0,\infty ]$.  Accordingly.
\begin{equation}
 z = -\{ln(random number from [0,1]\}/x
\end{equation} as has been emphasized by Gillespie)
\cite{Gillespie}.  The variable $z$ obviously represents the
time spent in the successor state.

During a single run of the simulation the fluctuations among
the 8 possible states are large, making it difficult to ``eyeball'' an average time behavior to compare with the
X(t), Y(t), ... functions given by the solution to the
master equation. The conventional method is to use ensemble
averaging \cite{Newman}, by running the
simulation using the same initial state but with different
random number seeds. The states as a function of time are
combined and sorted into small time bins for which the
average values of X(t), Y(t),... are easily calculated. Typically
results from 50 to 100 runs are sufficient to obtain values
in which the fluctuations are sufficiently small (i.e.
standard deviation values for X(t) of 0.0806 and 0.0672
respectively) for reasonable comparisons to the master
equation solutions. Both the master eequation relaxation times and asymptotic
values are well reproduced by the simulation. Plots of these are
available at the web site.

	For a single interactive display run of the simulation
which shows the 3-spin state as a function of time, we also
show a plot of the X(t) and Y(t) function using a different
averaging technique to give the user an indication of how
the system relaxes to equilibrium. The algorithm
accumulates, at each instant,  the time spent in each of the states from time
zero, thereby estimating the probabilities of each state as a
function of time. This method reproduces the general
behavior of the X(t), Y(t), and asymptotic values.

	\begin{center} \textbf{ A Python program to simulate the
behavior of a 3 spin system using a Stochastic Monte Carlo
Approach:} \end{center}

	 \begin{verbatim} #!/usr/bin/python # this is a
code to do a stochastic simulation of the dynamic time
evolution of a state consisting of # 3 spin 1/2 objects
based on a state transition diagram which specifies the
available transitions # among the 8 possible states and the
forward and backward transition rates. # Jack Uretsky and Ed
May 2011 # this is glauber3spinSimulation3.py # this is
revised to use exponential distribution in the A_ functions
# to calculate the time of the next transition based on the
rate local to the # A_ function # A graphical display (based
on Tk functions) of the state and the X and Y functions as a
function # of the iteration step. The display updates in the
time determined by the simulation

	import sys import time import random import math

	debuglevel = 0

	def expon(x) : z = random.random() expValue =
-math.log(z)/x return expValue

	if len(sys.argv) >= 1: print print "Usage:
view3spinSimulation3.py  stateid gamma numberOfIterations
seed" print "For example: view3spinSimulation3.py 1 0.25 325
1379" time.sleep(2)   # sleep for 2 seconds then continue
with the simulation using default configuration #sys.exit()
graphicsmode = "Tkinter" # view3spinSimulation3.py [ Tkinter
] stateid gamma numberOfIterations runN seed #
view3spinSimulation3.py             stateid gamma
numberOfIterations seed # 1                                 
 2       3     4                  5 # decode the command
arguement list to get initial parameters k=0 for arg in
sys.argv: #print "Command line",arg,k k=k+1 graphicsmode =
"Tkinter" from Tkinter import *

	initialstateid = 1      # all spins Up if len(sys.argv)
>= 2 : initialstateid = int(sys.argv[1]) initialgamma = 0.25
if len(sys.argv) >= 3 : initialgamma = float(sys.argv[2])
numberOfIter = 325      # fills the iteration length of the
display
if len(sys.argv) >= 4 : numberOfIter = int(sys.argv[3])
initialrunN = 1
initialseed = 137
if len(sys.argv) >= 5 : initialseed = int(sys.argv[4])

if initialgamma <=-1.0 :
        print "Unphysical value for gamma...Exit"
        exit()

if initialgamma >=+1.0 :
        print "Unphysical value for gamma...Exit"
        exit()

print
print "User requested graphicsmode=",graphicsmode,"initialstateid=",initialstateid
print "gamma=",initialgamma,"numberOfIter=",numberOfIter, "initialseed=",initialseed
print

from sys import exit
import time
import random
from math import *
global gamma
gamma = initialgamma
runN=initialrunN
# setup structures to communicate with A_ functions
global RET
RET = [0,0,0]   # will contain the triple state, rate, time given by the A_ functions
global state
global r
global N

# setup a dictionary to store state spin patterns for 3 spins 0=spin down 1=spin up
def bits(state) :
        s=''
        #t={1:'111', 2:'110', 3:'101', 4:'100', 5:'011', 6:'010', 7:'001', 8:'000'} #using appendix A
        t={1:'111', 2:'011', 3:'101', 4:'110', 5:'100', 6:'010', 7:'001', 8:'000'} # jacks 21oct10 email
        s=t[state]
        return s

# The A_ functions change the state at time t based on state diagram
def A_1():
    state = 1                   # current state
    r = 1 - gamma
    f = random.randrange(3)
    g = f+2  #destination
    t = expon( 3*r )
    RET = [g,r,t]               # g is new state after time t
    return RET

def A_2():
        state = 2
        global r2
        r2 = 1. + gamma
        f = random.random()
        if f > 2./(3. + gamma):
                g = 1
        elif f > 1./(3. + gamma):
                g,r2 = 7,1
        else:
                g,r2 = 6,1
        t =  expon(r2+2.0)
        RET = [g,r2,t]
        return RET
        
def A_3():
        state = 3
        global r3
        r3 = 1. + gamma
        f = random.random()
        if f > 2./(3. + gamma):
                g = 1
        elif f > 1./(3. + gamma):
                g,r3 = 7,1
        else:
                g,r3 = 5,1
        t = expon(r3+2.0)
        RET = [g,r3,t]
        return RET
        
def A_4():
        state = 4
        global r4
        r4 = 1. + gamma
        f = random.random()
        if f > 2./(3. + gamma):
                g = 1
        elif f > 1./(3. + gamma):
                g,r4 = 5,1      
        else:
                g,r4 = 6,1
        t = expon(r4+2.0)
        RET = [g,r4,t]
        return RET
        
def A_5():
        state = 5
        global r5
        r5 = 1. + gamma
        f = random.random()
        if f > 2./(3. + gamma):
                g = 8
        elif f > 1./(3. + gamma):
                g,r5 = 4,1
        else:

                g,r5 = 3,1
        t = expon(r5+2.0)
        RET = [g,r5,t]
        return RET
        
def A_6():
        state = 6
        global r6
        r6 = 1. + gamma
        f = random.random()
        if f > 2./(3. + gamma):
                g = 8
        elif f > 1./(3. + gamma):
                g,r6 = 4,1
        else:
                g,r6 = 2,1
        t = expon(r6+2.0)
        RET = [g,r6,t]
        return RET
        
def A_7():
        state = 7
        global r7
        r7 = 1. + gamma
        f = random.random()
        if f > 2./(3. + gamma):
                g = 8
        elif f > 1./(3. + gamma):
                g,r7 = 3,1
        else:
                g,r7 = 2,1
        t = expon(r7+2.0)
        RET = [g,r7,t]

        return RET
        
def A_8():
    state = 8
    global r8
    r8 = 1. - gamma
    f = random.randrange(3)
    g = f + 5
    t = expon(3*r8)
    RET = [g, r8, t]
    return RET

#dictionaries
Adestin = {1: A_1, 2 : A_2, 3 : A_3, 4: A_4,5 : A_5, 6 : A_6, 7: A_7, 8 : A_8}
apics = {1: 'IMAGES/a_1.jpg', 2: 'IMAGES/a_2.jpg', 3 : 'IMAGES/a_3.jpg', 4: 'IMAGES/a_4.jpg',
         5: 'IMAGES/a_5.jpg', 6: 'IMAGES/a_6.jpg', 7 : 'IMAGES/a_7.jpg', 8 : 'IMAGES/a_8.jpg'}
state = initialstateid
print "Beginning state=",state
iseed=initialseed
random.seed(iseed)
# open a data output file to store simulation data for this run
openfile=open('glauber3spinSimulation.data','w')
# open a data output file to store configuration parameters for this run, used for plotting
openfileconf=open('glauber3spinSimulation.conf','w')

back1 = apics[state]    # setup the initial state

if graphicsmode == "Tkinter" :
        # Setup Tkinter graphics
        # create the canvas, size in pixels
        canvas = Canvas(width = 700, height = 750, bg = 'yellow')
        # pack the canvas into a frame/form
        canvas.pack(expand = YES, fill = BOTH)
        # create the X axis (ie iteration count)
        for i in range(50,700)  :
                canvas.create_text(i,625,fill='black',text='-')
        asymX = (3*(1-gamma))/(2*(2-gamma))
        j=625-100*asymX
        for i in range(650,700) :
                canvas.create_text(i,j,fill='black',text='.')
        canvas.create_text(600,625,fill='black',text='iteration')
        canvas.create_text(600,525,fill='red',text='X() simulation')
        canvas.create_text(600,545,fill='blue',text='Y() simulation')
        # create the Y axis at x=50
        for i in range(525,725) :
                canvas.create_text(50,i,fill='black',text='.')
        # create the Y axis tick marks
        for i in range(45,55) :
                canvas.create_text(i,525,fill='black',text='.')
                canvas.create_text(i,625,fill='black',text='-')
                canvas.create_text(i,725,fill='black',text='.')
        canvas.create_text(30,525,fill='black',text=' 1.0')
        canvas.create_text(30,625,fill='black',text=' 0.0')
        canvas.create_text(30,725,fill='black',text='-1.0')
        canvas.update()

k=0
vPlusCnt=0.
vMinusCnt=0.
xPlusCnt=0.
xMinusCnt=0.
pXsum=0.0
pX=0.0
pY=0.0
arbTime = 0.
xAtTime =0.
xAtZero=0.0
statePop=[]
for i in range(8) :
        statePop.append(0)

# write configuration data information
openfile.write('# iseed = '+str(iseed)+' init state = '+str(state)+' gamma = '+str(initialgamma) + '\n')
openfile.write('# numberOfIterations = '+str(numberOfIter) + '\n')

# write out header
openfile.write('# rowIndex'+' '+'state'+' '+'arbTime'+' '+'runN'+' '+'pX'+' '+'pY'+'\n') 
# write out starting state information
openfile.write(str(k)+' '+str(state)+' '+str(arbTime)+' '+str(runN)+'\n')

while k <= numberOfIter :
    startTime = time.time()
    oldState=state
    k = k+1
    back1 = apics[state]
    RET = Adestin[state]()               # this executes the appropriate A_ function to get a new state
    destin = RET[0]
    r = RET[1]
    state = destin
    if graphicsmode == "Tkinter" :
        # display the state using Tkinter graphics
        filename = 'IMAGES/a_'+str(state)+'.gif'
        filename = 'IMAGES/a_'+str(oldState)+'.gif'
        gifimage = PhotoImage(file = filename)
        canvas.create_image(20, 10, image = gifimage, anchor = NW)
        j= 625-100*pX           # calc a x,y position to plot pX value
        k2 = 2*k+50
        canvas.create_text(k2,j,fill='red',text='x')
        j= 625-100*pY           # calc a x,y position to plot pY value
        canvas.create_text(k2,j,fill='blue',text='+')
        canvas.update()
    # get time generated in the A_ functions
    rest = RET[2]
    arbTime=arbTime+rest
    statePop[oldState-1] += rest                # update a time weighted state 
    state1Prob = float(statePop[0])/arbTime
    #
    #t={1:'111', 2:'011', 3:'101', 4:'110', 5:'100', 6:'010', 7:'001', 8:'000'} # jacks 21oct10 email
    #if (state==1) : vPlusCnt=vPlusCnt+1
    #if (state==8) : vMinusCnt=vMinusCnt+1
    #if ( state==2 or state==3 or state==4 ) : xPlusCnt=xPlusCnt+1
    #if ( state==5 or state==6 or state==7 ) : xMinusCnt=xMinusCnt+1
    # calculate various time weighted average for various X,Y,U,V functions at current time
    vPlusCnt = statePop[0]
    vMinusCnt= statePop[7]
    xPlusCnt = statePop[1]+statePop[2]+statePop[3]
    xMinusCnt= statePop[4]+statePop[5]+statePop[6]
    pVplus = float(vPlusCnt)/arbTime
    pVminus = float(vMinusCnt)/arbTime
    pV= pVplus + pVminus
    pXplus = float(xPlusCnt)/arbTime
    pXminus = float(xMinusCnt)/arbTime
    pX = pXplus + pXminus
    pU = pVplus - pVminus

    pY = pXplus - pXminus
    #pXsum = pXsum + oldState*rest      #?
    #pXave = pXsum/arbTime
    xAtTime=xAtZero*exp(-2*(2-gamma)*arbTime) +(3*(1-gamma)*(1.0 - exp(-2*(2-gamma)*arbTime))/(2*(2-gamma)))
    #sys.stdout.write('\r'+'at iteration='+str(k)+' state= '+bits(oldState)+' pX='+str(pX)+' time='+ str(arbTime))
    print "\rAt iteration=%d state=%s pX=%6.4f time=%6.3f" % (k,bits(oldState),pX,arbTime),
    sys.stdout.flush()
    openfile.write(str(k)+' '+str(state)+' '+str(arbTime)+' '+str(runN)+' '+str(pX)+' '+str(pY)+'\n')
    time.sleep(rest)

print "\n"
vPlusCnt = statePop[0]
vMinusCnt= statePop[7]
xPlusCnt = statePop[1]+statePop[2]+statePop[3]
xMinusCnt= statePop[4]+statePop[5]+statePop[6]
pVplus = float(vPlusCnt)/arbTime

pVminus = float(vMinusCnt)/arbTime
pV= pVplus + pVminus
pXplus = float(xPlusCnt)/arbTime
pXminus = float(xMinusCnt)/arbTime
pX = pXplus + pXminus
pU = pVplus - pVminus
pY = pXplus - pXminus
print "At end of iterations X=%5.3f Y=%5.3f V=%5.3f U=%5.3f" % (pX,pY,pV,pU)
if ( debuglevel > 0 ) :
   for i in range(8):
        print "%5.3f " % statePop[i-1],
   print
   for i in range(8):
        print "%5.3f " % (float(statePop[i-1])/arbTime),
   print

print
print "lastTime = %8.3f" % arbTime
print "gamma =",gamma
print "Running at single spin flip rate (flips/sec) = %5.3f" % (numberOfIter/arbTime)
print "initialseed =",initialseed
print "Master Equation predicition at gamma = ", gamma," and t--> infinity",
print "X=%5.3f" % ( (3*(1-gamma))/(2*(2-gamma)) )

openfileconf.write("lastTime = "+str(arbTime)+'\n')
openfileconf.write("gamma = "+str(gamma)+'\n')
openfileconf.write("initialseed = "+str(initialseed)+'\n')
print "end of program"

openfile.close()
openfileconf.close()
print '\a', '\a', '\a'  # ring terminal bell
time.sleep(10)
           
print "That's All, Folks!"
\end{verbatim}

\end{document}